\documentclass{article}
\usepackage{frascatiphys,here,graphicx,subfigure}
\begin{document}
\title{IS {\boldmath $\eta^\prime$} PARTIALLY MADE  OF GLUONIUM?
}
\author{
Rafel Escribano\\
{\em
Grup de F\'{\i}sica Te\`orica and IFAE,
Universitat Aut\`onoma de Barcelona,}\\
{\em
E-08193 Bellaterra (Barcelona), Spain}
}
\maketitle
\baselineskip=11.6pt
\begin{abstract}
A phenomenological analysis of radiative $V\to P\gamma$ and $P\to V\gamma$ decays is performed
in order to determine the gluonic content of the $\eta^\prime$ wave function.
Our result shows that there is no evidence for such a gluonium contribution,
$Z_{\eta^\prime}^2=0.04\pm 0.09$.
In terms of a mixing angle description this corresponds to $\phi_P=(41.4\pm 1.3)^\circ$ and
$|\phi_{\eta^\prime G}|=(12\pm 13)^\circ$.
\end{abstract}
\baselineskip=14pt
\section{Introduction}
\label{intro}
Is $\eta^\prime$ partially made of gluonium?
To answer this question we perform a phenomenological analysis of radiative
$V\to P\gamma$ and $P\to V\gamma$ decays,
with $V=\rho, K^\ast, \omega, \phi$ and $P=\pi, K, \eta, \eta^\prime$,
in order to determine the gluonic content of the $\eta^\prime$ wave function.
Similar analyses were driven in the seminal work by Rosner \cite{Rosner:1982ey},
where the allowed gluonic admixture in the $\eta^\prime$ could not be established due to the lack of data on $\phi\to\eta^\prime\gamma$, and, later on, by Kou who pointed out that the $\eta^\prime$ gluonic component might be as large as 26\% \cite{Kou:1999tt}.
More recently, the study by C.~E.~Thomas over a large number of different processes
concludes that while the data hint at a small gluonic component in the $\eta^\prime$,
the results depend sensitively on unknown form factors associated with exclusive dynamics
\cite{Thomas:2007uy}. 

From the experimental side,
the KLOE Collaboration, combining the new measurement of 
$R_\phi\equiv B(\phi\to\eta^\prime\gamma)/B(\phi\to\eta\gamma)$ with other constraints,
has estimated the gluonium content of the $\eta^\prime$ meson as
$Z_{\eta^\prime}^2=0.14\pm 0.04$ \cite{Ambrosino:2006gk}.
This new result contrasts with the former value $Z_{\eta^\prime}^2=0.06^{+0.09}_{-0.06}$,
which was compatible with zero and consistent with a gluonium fraction below 15\%
\cite{Aloisio:2002vm}.
The sole difference between the two analyses  is the inclusion in the amplitudes of
Ref.~\cite{Ambrosino:2006gk} of two extra parameters to deal with the overlap of the vector and pseudoscalar meson wave functions produced in the transitions $V\to P\gamma$ or $P\to V\gamma$,
a feature first introduced in Ref.~\cite{Bramon:2000fr}.
However, the new analysis of Ref.~\cite{Ambrosino:2006gk} uses the most recent experimental data taken from Ref.~\cite{Yao:2006px} in association with the values for the parameters related to the overlap which were obtained in Ref.~\cite{Bramon:2000fr} from a fit to available experimental data at that time.
Therefore, a reanalysis of this uncomfortable situation is needed before drawing definite conclusions on the gluon content of the $\eta^\prime$ meson.
This is the main motivation of the present work.
A more extensive version including a detailed analysis also for the case of the $\eta$,
the effects of considering the newest (not reported in the PDG) data,
a comparison with other approaches,
and a discussion of the $P\to\gamma\gamma$ decays within this context
can be found in Ref.~\cite{Escribano:2007cd}.

\section{Notation}
\label{notation}
We will work in a basis consisting of the states
 $|\eta_q\rangle\equiv\frac{1}{\sqrt{2}}|u\bar u+d\bar d\rangle$, $|\eta_s\rangle=|s\bar s\rangle$
and $|G\rangle\equiv|\mbox{gluonium}\rangle$.
The physical states $\eta$ and $\eta^\prime$ are assumed to be linear combinations of these:
\begin{equation}
\label{mixingP}
\begin{array}{rcl}
|\eta\rangle &=& X_\eta|\eta_q\rangle+Y_\eta|\eta_s\rangle+Z_\eta|G\rangle\ ,\\[1ex]
|\eta^\prime\rangle &=&
X_{\eta^\prime}|\eta_q\rangle+Y_{\eta^\prime}|\eta_s\rangle+Z_{\eta^\prime}|G\rangle\ ,
\end{array}
\end{equation}
with $X_{\eta(\eta^\prime)}^2+Y_{\eta(\eta^\prime)}^2+Z_{\eta(\eta^\prime)}^2=1$
and thus $X_{\eta(\eta^\prime)}^2+Y_{\eta(\eta^\prime)}^2\leq 1$.
A significant gluonic admixture in a state is possible only if
$Z_{\eta(\eta^\prime)}^2=1-X_{\eta(\eta^\prime)}^2-Y_{\eta(\eta^\prime)}^2>0$ \cite{Rosner:1982ey}.
This mixing scheme assumes isospin symmetry, \emph{i.e.~}no mixing with $\pi^0$,
and neglects other possible admixtures from $c\bar c$ states and/or radial excitations.
An interesting situation occurs when the gluonium content of the $\eta$ meson is assumed to vanish, $Z_{\eta}\equiv 0$.
In this particular case, the rotation between the physical states ($\eta$, $\eta^\prime$ and $\iota$) and the orthonormal mathematical states ($\eta_q$, $\eta_s$ and $G$) can be written in terms of two mixing angles, $\phi_P$ and $\phi_{\eta^\prime G}$,
which would correspond to 
\begin{equation}
\label{gluoniumetap}
\begin{array}{lll}
X_\eta=\cos\phi_P \ , & Y_\eta=-\sin\phi_P\ , & Z_\eta=0 \ ,\\[1ex]
X_{\eta^\prime}=\sin\phi_P\cos\phi_{\eta^\prime G} \ , &
Y_{\eta^\prime}=\cos\phi_P\cos\phi_{\eta^\prime G} \ , & Z_{\eta^\prime}=-\sin\phi_{\eta^\prime G} \ ,
\end{array}
\end{equation}
where $\phi_P$ is the $\eta$-$\eta^\prime$ mixing angle (in the quark-flavour basis)
in absence of gluonium, \emph{i.e.~}$\phi_{\eta G}=\phi_{\eta^\prime G}=0$.
It is related to its octet-singlet basis analog through
$\theta_P=\phi_P-\arctan\sqrt{2}\simeq\phi_P-54.7^\circ$.

\section{A model for {\boldmath $VP\gamma$ $M1$} transitions}
\label{model}
We will work in a conventional quark model where pseudoscalar and vector mesons are
simple quark-antiquark $S$-wave bound states with characteristic spatial extensions fixed by their respective quark-antiquark $P$ or $V$ wave functions.  
We take the good $SU(2)$ limit with $m_u = m_d \equiv\bar{m}$
and with identical spatial extension of wave functions within each $P$ and each $V$ isomultiplet.
$SU(3)$ will be broken in the usual manner taking constituent quark masses with
$m_s > \bar{m}$ but also, and this is a specific feature of our approach, 
allowing for different spatial extensions for each $P$ and $V$ isomultiplet.
Finally, we will consider that $VP\gamma$ transitions fully respect the usual OZI-rule. 

In our specific case of $VP\gamma$ $M1$ transitions, these generic statements translate 
into three characteristic ingredients of the model: 
{\it i)}
A $VP\gamma$ magnetic dipole transition proceeds via quark or antiquark spin-flip 
amplitudes proportional to $\mu_q=e_q/2m_q$. 
This effective magnetic moment breaks $SU(3)$ in a well defined way and distinguishes photon emission from strange or non-strange quarks via $m_s > \bar{m}$;
{\it ii)}
The spin-flip $V\leftrightarrow P$ conversion amplitude has then to be corrected by the 
relative overlap between the $P$ and $V$ wave functions;
{\it iii)}
Indeed, the OZI-rule reduces considerably the possible transitions and their respective 
$VP$ wave-function overlaps: $C_s$, $C_q$ and $C_\pi$ characterize the 
$\langle\eta_s|\phi_s\rangle$, 
$\langle\eta_q|\omega_q\rangle =\langle\eta_q|\rho\rangle$ and 
$\langle\pi|\omega_q\rangle =\langle\pi|\rho\rangle$ spatial overlaps, respectively.
Notice that distinction is made between the $|\pi\rangle$ and $|\eta_q\rangle$ 
spatial extension due to the gluon or $U(1)_A$ anomaly.

The relevant $VP\gamma$ couplings are written in terms of a
$g\equiv g_{\omega_q\pi\gamma}$ as
\begin{equation}
\label{etacouplings}
\begin{array}{c}
g_{\rho\eta^{(\prime)}\gamma}=g\,z_{q}\,X_\eta^{(\prime)}\ ,\\[2ex]
g_{\omega\eta^{(\prime)}\gamma}=
\frac{1}{3}g\left(z_q\,X_\eta^{(\prime)}\cos\phi_V+
2\frac{\bar{m}}{m_s}z_s\,Y_\eta^{(\prime)}\sin\phi_V\right)\ ,\\[2ex]
g_{\phi\eta^{(\prime)}\gamma}=
\frac{1}{3}g\left(z_q\,X_\eta^{(\prime)}\sin\phi_V-
2\frac{\bar{m}}{m_s}z_s\,Y_\eta^{(\prime)}\cos\phi_V\right)\ ,\\[2ex]
\end{array}
\end{equation}
where we have redefined $z_q\equiv C_q/C_\pi$ and $z_s\equiv C_s/C_\pi$.

\section{Data fitting}
\label{datafit}
We proceed to fit our theoretical expressions for the amplitudes 
comparing the available experimental information on 
$\Gamma (V\rightarrow P\gamma)$ and $\Gamma (P\rightarrow V\gamma)$
taken exclusively from Ref.~\cite{Yao:2006px}.
In the following, we leave the $z$'s free and allow for gluonium in the $\eta^\prime$ wave function only.
This will permit us to fix the gluonic content of the $\eta^\prime$ in a way identical to the experimental measurement by KLOE, that is, under the hypothesis of no gluonium in the $\eta$ wave function.
Unfortunately, a simultaneous fit of the $z$'s and the gluonic admixture in the
$\eta$ \emph{and} $\eta^\prime$ is not possible.
However, as a matter of comparison, we first consider the absence of gluonium in both mesons,
\emph{i.e.~}$\phi_{\eta G}=\phi_{\eta^\prime G}=0$.
The result of the fit gives $\chi^2/$d.o.f.=4.4/5 with
\begin{equation}
\label{zphiP}
\begin{array}{c}
g=0.72\pm 0.01\ \mbox{GeV$^{-1}$}\ ,\quad \phi_P=(41.5\pm 1.2)^\circ\ ,\quad
\phi_V=(3.2\pm 0.1)^\circ\ ,\\[2ex]
\frac{m_s}{\bar m}=1.24\pm 0.07\ ,\quad
z_q=0.86\pm 0.03\ ,\quad z_s=0.78\pm 0.05\ .
\end{array}
\end{equation}
If we fix the $z$'s to unity, the fit gets much worse ($\chi^2/$d.o.f.=45.9/8).
This shows that allowing for different overlaps of quark-antiquark wave functions and, in particular, for those coming from the gluon anomaly affecting only the $\eta$ and $\eta^\prime$ singlet component, is indeed relevant.

%
\begin{table}
\centering
\begin{tabular}{cccc}
\hline\\[-1.5ex]
Transition & $g_{VP\gamma}^{\rm exp}$(PDG) & 
$g_{VP\gamma}^{\rm th}$(Fit 1) & $g_{VP\gamma}^{\rm th}$(Fit 2)\\[1ex]
\hline\\[-1.5ex]
$\rho^0\rightarrow\eta\gamma$ &
$0.475\pm 0.024$ & $0.461\pm 0.019$ & $0.464\pm 0.030$\\[1ex]
$\eta^\prime\rightarrow\rho^0\gamma$ &
$0.41\pm 0.03$ & $0.41\pm 0.02$ & $0.40\pm 0.04$\\[1ex]
$\omega\rightarrow\eta\gamma$ &
$0.140\pm 0.007$ & $0.142\pm 0.007$ & $0.143\pm 0.010$\\[1ex]
$\eta^\prime\rightarrow\omega\gamma$ &
$0.139\pm 0.015$ & $0.149\pm 0.006$ & $0.146\pm 0.014$\\[1ex]
$\phi\rightarrow\eta\gamma$ &
$0.209\pm 0.002$ & $0.209\pm 0.018$ & $0.209\pm 0.013$\\[1ex]
$\phi\rightarrow\eta^\prime\gamma$ &
$0.22\pm 0.01$ & $0.22\pm 0.02$ & $0.22\pm 0.02$\\[1ex]
\hline
\end{tabular}
\caption{\it Comparison between the experimental values $g_{VP\gamma}^{\rm exp}$ (in GeV$^{-1}$)
for the transitions involving $\eta$ or $\eta^\prime$ taken from the PDG \cite{Yao:2006px}
and the corresponding predictions for $g_{VP\gamma}^{\rm th}$
from Eqs.~(\ref{zphiP}) ---Fit 1--- and (\ref{zphiPphietapG}) ---Fit 2---.}
\label{table1}
\end{table}
%
In Table \ref{table1}, we present a comparison between experimental data for the relevant
$VP\gamma$ transitions with $P=\eta,\eta^\prime$ and the corresponding theoretical predictions
(in absolute value) calculated from the fitted values in Eq.~(\ref{zphiP}).
The agreement is very good and all the predictions coincide with the experimental values within
$1\sigma$.

Now that we have performed a fit under the hypothesis of no gluonium we assume $\phi_{\eta G}=0$, \textit{i.e.}~$Z_\eta=0$,
and then proceed to fit the gluonic content of the $\eta^\prime$ wave function under this assumption.
The results of the new fit are\footnote{
There is a sign ambiguity in $\phi_{\eta^\prime G}$ that cannot be decided since this angle
enters into $X_{\eta^\prime}$ and $Y_{\eta^\prime}$ through a cosine.}
\begin{equation}
\label{zphiPphietapG}
\begin{array}{c}
g=0.72\pm 0.01\ \mbox{GeV$^{-1}$}\ ,\quad \frac{m_s}{\bar m}=1.24\pm 0.07\ ,\quad
\phi_V=(3.2\pm 0.1)^\circ\ ,\\[2ex]
\phi_P=(41.4\pm 1.3)^\circ\ ,\quad |\phi_{\eta^\prime G}|=(12\pm 13)^\circ\ ,\\[2ex]
z_q=0.86\pm 0.03\ ,\quad z_s=0.79\pm 0.05\ ,
\end{array}
\end{equation}
with $\chi^2/$d.o.f.=4.2/4.
The quality of the fit is similar to the one obtained assuming a vanishing gluonic admixture for both mesons ($\chi^2/$d.o.f.=4.4/5).
The result obtained for $\phi_{\eta^\prime G}$ suggests a very small amount of gluonium in the
$\eta^\prime$ wave function, 
$|\phi_{\eta^\prime G}|=(12\pm 13)^\circ$ or $Z_{\eta^\prime}^2=0.04\pm 0.09$.
This is the main result of our analysis.
Our values contrast with those reported by KLOE recently, $\phi_P=(39.7\pm 0.7)^\circ$ and
$|\phi_{\eta^\prime G}|=(22\pm 3)^\circ$ ---or $Z_{\eta^\prime}^2=0.14\pm 0.04$---
\cite{Ambrosino:2006gk}.
In Table \ref{table1}, we also include the theoretical predictions for the various transitions involving
$\eta$ or $\eta^\prime$ calculated from the fitted values in Eq.~(\ref{zphiPphietapG}).
As expected, there is no significant difference between the values obtained allowing for gluonium (Fit 2) or not (Fit 1) in the $\eta^\prime$ wave function.

%
\begin{figure}[t]
\centerline{\includegraphics[width=0.70\textwidth]{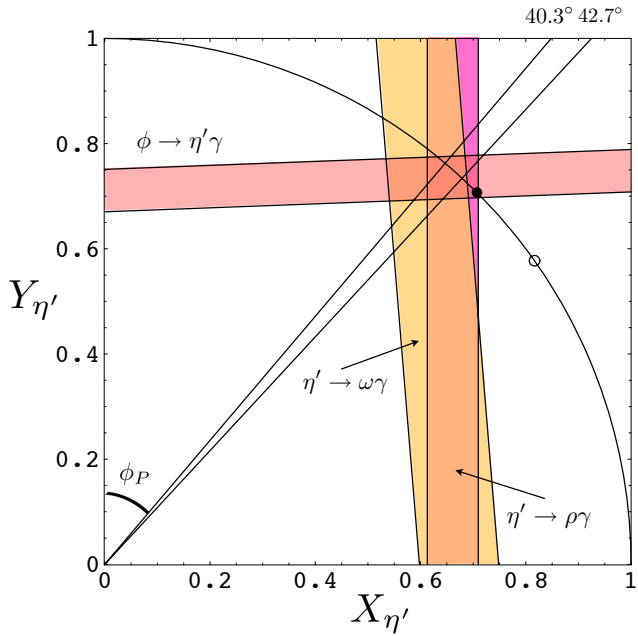}}  
\caption{Constraints on non-strange ($X_{\eta^\prime}$) and strange ($Y_{\eta^\prime}$) quarkonium mixing coefficients in the $\eta^\prime$.
The mixing solutions corresponding to the $\eta^\prime$ being a pure singlet
($X_{\eta^\prime}=\sqrt{2}Y_{\eta^\prime}=\frac{1}{\sqrt{3}}$) ---open circle--- and
($X_{\eta^\prime}=Y_{\eta^\prime}=\frac{1}{\sqrt{2}}$) ---closed circle--- are shown.
The vertical and inclined bands are the regions for $X_{\eta^\prime}$ and $Y_{\eta^\prime}$ allowed by the experimental couplings of the $\eta^\prime\to(\rho,\omega)\gamma$ and
$\phi\to\eta^\prime\gamma$ transitions.}
\label{plotetap}
\end{figure}
%
Our main results can also be displayed graphically following
Refs.~\cite{Rosner:1982ey,Kou:1999tt,Ambrosino:2006gk}.
In Fig.~\ref{plotetap}, we plot the regions for the $X_{\eta^\prime}$ and $Y_{\eta^\prime}$ parameters which are allowed by the experimental couplings of the
$\eta^\prime\to\rho\gamma$, $\eta^\prime\to\omega\gamma$ and $\phi\to\eta^\prime\gamma$ transitions (see Table \ref{table1}).
The limits of the bands are given at 68\% CL or $1\sigma$.
The remaining parameters are taken from Eq.~(\ref{zphiPphietapG}).
In addition to the bands, we have also plotted the circular boundary denoting the constraint 
$X_{\eta^\prime}^2+Y_{\eta^\prime}^2\leq 1$ as well as the favoured region for the
$\eta$-$\eta^\prime$ mixing angle assuming the \emph{absence of gluonium},
$40.3^\circ\leq\phi_P\leq 42.7^\circ$, obtained at $1\sigma$
from the corresponding fitted value in Eq.~(\ref{zphiP}).
There exists an intersection region of the three bands inside and on the circumference.
As most of this region is interior but close to the circular boundary it may well indicate a small but non necessarily zero gluonic content of the $\eta^\prime$.
Indeed, we have found $Z_{\eta^\prime}^2=0.04\pm 0.09$ (or $|Z_{\eta^\prime}|=0.2\pm 0.2$) or using the angular description $|\phi_{\eta^\prime G}|=(12\pm 13)^\circ$.
The size of the error is precisely what prevent us from drawing a definite conclusion concerning the amount of gluonium in the $\eta^\prime$ wave function.
More refined experimental data, particularly for the $\phi\to\eta^\prime\gamma$ channel, will contribute decisively to clarify this issue (see below).
Clearly, the inclusion of this process is of major importance for the determination of the gluonic admixture in the $\eta^\prime$, as observed for the first time in Ref.~\cite{Rosner:1982ey}.
In the present analysis, the ``democratic'' mixing solution is excluded at the $1\sigma$ level whereas the singlet solution is clearly excluded.

%
\begin{figure}[t]
\centerline{\includegraphics[width=0.65\textwidth]{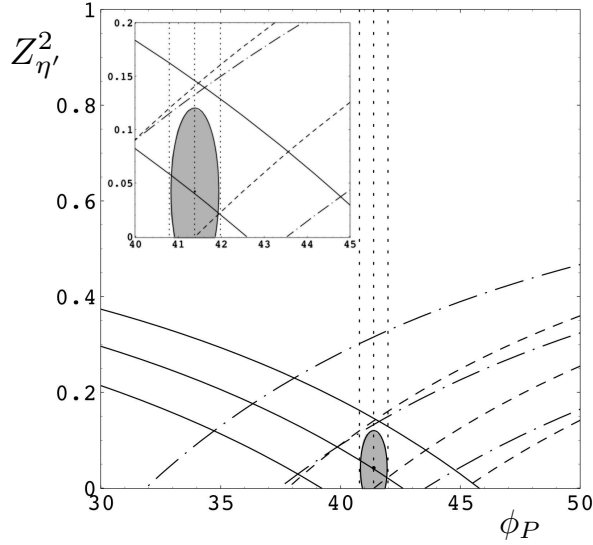}}  
\caption{The gray ellipse in the $(\phi_P,Z_{\eta^\prime}^2)$ plane corresponds to the allowed region at 68\% CL of the solution in Eq.~(\ref{zphiPphietapG}) assuming the presence of gluonium.
The different bands are the regions for $\phi_P$ and $Z_{\eta^\prime}^2$ allowed by the experimental couplings of the $\eta^\prime\to\rho\gamma$ (dashed line),
$\eta^\prime\to\omega\gamma$ (dot-dashed line), $\phi\to\eta\gamma$ (dotted line), and
$\phi\to\eta^\prime\gamma$ (solid line) transitions in Table \ref{table1}.}
\label{italians}
\end{figure}
%
To make our bounds more graphical, we follow Ref.~\cite{Ambrosino:2006gk} and
plot in Fig.~\ref{italians} the constraints from $\eta^\prime\to(\rho,\omega)\gamma$ and
$\phi\to(\eta,\eta^\prime)\gamma$ in the $(\phi_P,Z_{\eta^\prime}^2)$ plane together with the 68\% CL allowed region for gluonium as obtained from Eq.~(\ref{zphiPphietapG}).
The point corresponding to the preferred solution, $(\phi_P,Z_{\eta^\prime}^2)=(41.4^\circ,0.04)$,
is also shown.
The allowed region is very constrained in the $\phi_P$ axis by the experimental value of the
$g_{\phi\eta\gamma}$ coupling, whose vertical band denotes its non dependence on
$Z_{\eta^\prime}^2$.
The other three bands, all dependent on $\phi_P$ and $Z_{\eta^\prime}^2$, constrain the amount of gluonium down to a value compatible with zero at $1\sigma$.

\section{Summary and conclusions}
\label{conclusions}
In this work we have performed a phenomenological analysis of radiative $V\to P\gamma$ and
$P\to V\gamma$ decays with the purpose of determining the gluon content of the $\eta^\prime$ meson.
The present approach is based on a conventional $SU(3)$ quark model supplemented with two sources of $SU(3)$ breaking, the use of constituent quark masses with $m_s>\bar m$ and the different spatial extensions for each $P$ or $V$ isomultiplet which induce different overlaps between the $P$ and $V$ wave functions.
The use of these different overlapping parameters ---a specific feature of our analysis---
is shown to be of primary importance in order to reach a good agreement.

Our conclusions are the following.
First, accepting the absence of gluonium for the $\eta$ meson, the current experimental data on
$VP\gamma$ transitions indicate within our model a negligible gluonic content for the
$\eta^\prime$ meson, $Z_{\eta^\prime}^2=0.04\pm 0.09$.
Second, this gluonic content of the 
$\eta^\prime$ wave function amounts to $|\phi_{\eta^\prime G}|=(12\pm 13)^\circ$ and the
$\eta$-$\eta^\prime$ mixing angle is found to be $\phi_P=(41.4\pm 1.3)^\circ$.
Third, imposing the absence of gluonium for both mesons one finds $\phi_P=(41.5\pm 1.2)^\circ$,
in agreement with the former result.
Finally, we would like to stress that more refined experimental data, particularly for the
$\phi\to\eta^\prime\gamma$ channel, will contribute decisively to clarify this issue.

\begin{thebibliography}{99}
\bibitem{Rosner:1982ey}
  J.~L.~Rosner,
  Phys.\ Rev.\  D {\bf 27} (1983) 1101.

\bibitem{Kou:1999tt}
  E.~Kou,
  Phys.\ Rev.\  D {\bf 63} (2001) 054027
  [arXiv:hep-ph/9908214].

\bibitem{Thomas:2007uy}
  C.~E.~Thomas,
  JHEP {\bf 0710} (2007) 026
  [arXiv:0705.1500 [hep-ph]].

\bibitem{Ambrosino:2006gk}
  F.~Ambrosino {\it et al.}  [KLOE Collaboration],
  arXiv:hep-ex/0612029.

\bibitem{Aloisio:2002vm}
  A.~Aloisio {\it et al.}  [KLOE Collaboration],
  Phys.\ Lett.\  B {\bf 541} (2002) 45
  [arXiv:hep-ex/0206010].

\bibitem{Bramon:2000fr}
  A.~Bramon, R.~Escribano and M.~D.~Scadron,
  Phys.\ Lett.\  B {\bf 503} (2001) 271
  [arXiv:hep-ph/0012049].

\bibitem{Yao:2006px}
  W.~M.~Yao {\it et al.}  [Particle Data Group],
  J.\ Phys.\ G {\bf 33} (2006) 1.

\bibitem{Escribano:2007cd}
  R.~Escribano and J.~Nadal,
  JHEP {\bf 0705} (2007) 006
  [arXiv:hep-ph/0703187].
\end{thebibliography}
\end{document}